\documentclass{elsart}
\journal{Image and Vision Computing}
\usepackage{graphicx,color}
\usepackage{multirow,bigstrut}
\graphicspath{{figures/}}
\usepackage{amssymb,amsmath,bm}
\usepackage{cite,url}
\newcommand\mymatrix[1]{\bm{\mathrm{#1}}}
\definecolor{dgreen}{rgb}{0,.6,0}
\usepackage[normalem]{ulem}
\begin{document}
\newlength\imgwidth
\setlength\imgwidth{0.45\columnwidth}
\begin{frontmatter}
\title{Cryptanalysis of an Image Encryption Scheme Based on a Compound Chaotic Sequence}
\author[hk-cityu]{Chengqing Li\corauthref{corr}},
\author[germany]{Shujun Li\corauthref{corr}},
\author[hk-cityu]{Guanrong Chen} and
\author[germany]{Wolfgang A. Halang}
\corauth[corr]{Corresponding authors: Chengqing Li
(swiftsheep@hotmail.com), Shujun Li (http://www.hooklee.com).}
\address[hk-cityu]{Department of Electronic Engineering, City University of Hong Kong,
83 Tat Chee Avenue, Kowloon Tong, Hong Kong SAR, China}
\address[germany]{FernUniversit\"at in Hagen, Lehrstuhl f\"ur Informationstechnik,
58084 Hagen, Germany}

\begin{abstract}
Recently, an image encryption scheme based on a compound chaotic
sequence was proposed. In this paper, the security of the scheme is
studied and the following problems are found: (1) a differential
chosen-plaintext attack can break the scheme with only three chosen
plain-images; (2) there is a number of weak keys and some equivalent
keys for encryption; (3) the scheme is not sensitive to the changes
of plain-images; and (4) the compound chaotic sequence does not work
as a good random number resource.
\end{abstract}

\end{frontmatter}

\section{Introduction}

Security of multimedia data is receiving more and more attention due
to the widespread transmission over various communication networks.
It has been noticed that the traditional text encryption schemes
fail to safely protect multimedia data due to some special
properties of these data and some specific requirements of
multimedia processing systems, such as bulky size and strong
redundancy of uncompressed data. Therefore, designing good image
encryption schemes has become a focal research topic since the early
1990s. {Inspired by} the subtle similarity between chaos and
cryptography, a large number of chaos-based image encryption schemes
have been proposed
\cite{Yen:CKBA:ISCAS2000,Chen&Yen:RCES:JSA2003,Chen-Guo-Huang-Yen:TDCEA:EURASIP2003,
YaobinMao:CSF2004,Pareek:ImageEncrypt:IVC2006,Li:PermutationAnalysis:2004}.
Unfortunately, many of these schemes have been found insecure,
especially against known and/or chosen-plaintext attacks
\cite{Li:AttackTDCEA2005,Li-Zheng:CKBA:ISCAS2002,
Li:AttackingRCES2004,Li:AttackingIVC2007}. For a recent survey of
state-of-the-art image encryption schemes, the reader is referred to
\cite{ShujunLi:ChaosImageVideoEncryption:Handbook2004}. Some general
rules about evaluating the security of chaos-based cryptosystems can
be found in \cite{AlvarezLi:Rules:IJBC2006}.

Recently, an image encryption scheme based on a compound chaotic
sequence was proposed in \cite{Tong:ImageCipher:IVC07}. This scheme
includes two procedures: substitutions of pixel values with XOR
operations, and circular shift position permutations of rows and
columns. The XOR substitutions are controlled by a compound
pseudo-random number sequence generated from two correlated chaotic
maps. And the row and column circular shift permutations are
determined by the two chaotic maps, respectively. This paper studies
the security of the image encryption scheme and reports the
following findings:
\begin{enumerate}
\item the scheme can be broken by using only three chosen plain-images;
\item there exist some weak keys and equivalent keys;
\item the scheme is not sufficiently sensitive to the changes of
plain-images; and
\item the compound chaotic sequence is not random enough to be used
for encryption.
\end{enumerate}

This paper is organized as follows. In the next section the image
encryption scheme under study is briefly introduced. Then, in
Section~\ref{sec:SecurityProblem}, some security problems of the
scheme are discussed. A differential chosen plain-image attack is
introduced in Section~\ref{sec:Cryptanalysis} with some experimental
results reported. Finally, some conclusions are given in
Section~\ref{sec:conclusion}.

\section{The image encryption scheme under study}

Although not explicitly mentioned, the image encryption scheme was
specifically tailored to 24-bit RGB true-color images. However, the
algorithm itself is actually independent of the plain-image's
structure and can be used to encrypt any 2-D byte array. Therefore,
in this cryptanalytic paper, it is assumed that the plain-image is
an $M\times N$ (width$\times$height) 8-bit gray-scale image. In
other words, to encrypt a 24-bit RGB true-color image, one only
needs to consider the true-color image as a $3M\times N$ 8-bit
gray-scale image, and then perform the encryption procedure.

Denoting the plain-image by $\mymatrix{I}=\{I(i,j)\}_{1\leq i\leq M
\atop 1\leq j\leq N}$ and the corresponding cipher-image by
$\mymatrix{I}'=\{I'(i,j)\}_{1\leq i\leq M \atop 1\leq j\leq N}$, the
image encryption scheme proposed in \cite{Tong:ImageCipher:IVC07}
can be described as follows\footnote{To make the presentation more
concise and complete, some notations in the original paper are
modified, and some missed details about the encryption procedure are
supplied here.}.
\begin{itemize}
\item The \textit{secret key} includes two floating-point numbers of
precision $10^{-14}$ $x_0, y_0\in[-1,1]$, which are the initial
states of the following two chaotic maps: $f_0(x)=8x^4-8x^2+1$ and
$f_1(y)=4y^3-3y$.
\item The \textit{initialization procedure} includes generation of three
pseudo-random integer sequences.
\begin{enumerate}
\item \textit{Pseudo-random sequence $\{S_1(k)\}_{k=1}^{MN}$ for XOR
substitution of pixel values}

Starting from $k_0=k_1=0$, iterate the following compound chaotic
map for $MN$ times to construct a compound chaotic sequence
$\{z_k\}_{k=1}^{MN}$:
\begin{equation}
z_{k_0+k_1+1}=
\begin{cases}
x_{k_0+1}=f_0(x_{k_0}), & \mbox{if } (x_{k_0}+y_{k_1})<0,\\
y_{k_1+1}=f_1(y_{k_1}), & \mbox{if } (x_{k_0}+y_{k_1})\geq 0.
\end{cases}\label{eq:compoundmap}
\end{equation}
For each iteration of Eq.~(\ref{eq:compoundmap}), update $k_0$ with
$k_0+1$ if the first condition is satisfied, and update $k_1$ with
$k_1+1$ otherwise.

Then, an integer sequence $\{S_1(k)\}_{k=1}^{MN}$ is obtained from
$\{z_k\}_{k=1}^{MN}$ as
\begin{equation}
S_1(k)=\begin{cases}%
\left\lfloor \frac{1+z_k}{2}\cdot 256\right\rfloor, & \mbox{if }z_k\in[-1,1),\\
255, & \mbox{if }z_k=1,
\end{cases}
\end{equation}
where $\lfloor a \rfloor$ denotes the greatest integer that is not
greater than $a$.

\item \textit{Pseudo-random sequence $\{S_2(j)\}_{j=1}^N$ for circular
shift operations of rows}

Iterate $f_0$ from $x_{k_0}$ for $N$ more times to obtain a chaotic
sequence $\{x_{k_0+j}\}_{j=1}^N$, and then transform it into
$\{S_2(j)\}_{j=1}^N$ by
\[
S_2(j)=\begin{cases}%
\left\lfloor \frac{1+x_{k_0+j}}{2}\cdot M\right\rfloor, & \mbox{if }x_{k_0+j}\in[-1,1),\\
M-1, & \mbox{if }x_{k_0+j}=1.
\end{cases}
\]

\item \textit{Pseudo-random sequence $\{S_3(i)\}_{i=1}^M$ for circular
shift operations of columns}

Iterate $f_1$ from $y_{k_1}$ for $M$ more times to obtain a chaotic
sequence $\{y_{k_1+i}\}_{i=1}^M$, and then transform it into
$\{S_3(i)\}_{i=1}^M$ by
\[
S_3(i)=\begin{cases}%
\left\lfloor \frac{1+y_{k_1+i}}{2}\cdot N\right\rfloor, & \mbox{if }y_{k_1+i}\in[-1,1),\\
N-1, & \mbox{if }y_{k_1+i}=1.
\end{cases}
\]
\end{enumerate}
\item The \textit{encryption procedure} includes an XOR substitution part
and two permutation parts.
\begin{enumerate}
\item \textit{XOR substitution part}

Taking $\mymatrix{I}$ as input, an intermediate image
$\mymatrix{I}^*=\{I^*(i,j)\}_{1\leq i\leq M \atop 1\leq j\leq N}$ is
obtained as
\begin{equation}
I^*(i,j)=I(i,j)\oplus S_1((j-1)\cdot M+i),
\end{equation}
where $\oplus$ denotes the bitwise XOR operation.

\item \textit{Permutation part -- horizontal circular shift operations}

Taking $\mymatrix{I}^*$ as input, a new intermediate image
$\mymatrix{I}^{**}=\{I^{**}(i,j)\}_{1\leq i\leq M \atop 1\leq j\leq
N}$ is obtained by performing the following horizontal circular
shift operations\footnote{In \cite{Tong:ImageCipher:IVC07}, the
authors did not explain in which direction the circular shift
operations are performed. Since the direction is independent of the
scheme's security, here it is assumed that the operations are
carried out towards larger indices. The same assumption is made for
vertical circular shift operations.}:
\begin{equation}
I^{**}(i,j)=I^*((i-S_2(j))\bmod M,j).
\end{equation}

\item \textit{Permutation part -- vertical circular shift operations}

Taking $\mymatrix{I}^{**}$ as input, the cipher-image
$\mymatrix{I}'$ is obtained by performing the following vertical
circular shift operations:
\begin{equation}
I'(i,j)=I^{**}(i,(j-S_3(i))\bmod N).
\end{equation}
\end{enumerate}
Combining the above three operations, the encryption procedure can
be represented in the following compact form:
\begin{equation}\label{eq:fullencryption}
I'(i,j)=I(i^*,j^*)\oplus S_1((j^*-1)\cdot M+i^*),
\end{equation}
where $j^*=(j-S_3(i))\bmod N$ and $i^*=(i-S_2(j^*))\bmod M$.

\item The \textit{decryption procedure} is the reversion of the above
(after finishing the same initialization process) and can be
described as
\begin{equation}
I(i,j)=I'(i^*,j^*)\oplus S_1((j-1)\cdot M+i),
\end{equation}
where $i^*=(i+S_2(j))\bmod M$ and $j^*=(j+S_3(i^*))\bmod N$.
\end{itemize}

\section{Some security problems}\label{sec:SecurityProblem}

\subsection{Insufficient randomness of the compound chaotic sequence}
\label{ssec:Randomness}

In \cite[Sec. 4.3]{Tong:ImageCipher:IVC07}, the authors claim that
the randomness of the generated chaotic sequences has been verified
by employing the four random tests defined in FIPS PUB~140-2
\cite{FIPS1402:2002}. Here, it is noticed that what they actually
refer to is an intermediate edition of FIPS PUB 140-2 (updated in
October 2001), which has been superseded in December 2002, and as a
result all the four random tests have been removed from the
publication (see Change Notices~1 and 2, pp.~54--58 in
\cite{MOV:CyrptographyHandbook1996}).\footnote{In
\cite{Tong:ImageCipher:IVC07}, the authors cite
\cite{MOV:CyrptographyHandbook1996} as the source of FIPS PUB 140-2.
However, \cite{MOV:CyrptographyHandbook1996} only contains an
introduction to FIPS PUB 140-1 (the first edition of FIPS PUB 140)
\cite{FIPS1401:1994}. By comparing the required intervals shown in
Table~2 of \cite{Tong:ImageCipher:IVC07} with those published in
different editions of FIPS PUB 140, we finally concluded that FIPS
PUB 140-2 (Change 1) was the one used by the authors of
\cite{Tong:ImageCipher:IVC07}.}

Even for the four random tests defined in the intermediate edition
of FIPS PUB~140-2, the randomness of the chaotic sequences is still
questionable due to the following two facts:
\begin{enumerate}
\item Only the experimental result about one random sequence generated
from the key $(x_0, y_0)=(0.32145645647836, 0.48124356788345)$ is
shown in \cite{Tong:ImageCipher:IVC07}. However, to study the
randomness of a random number resource, a sufficiently large number
of samples should be tested.
\item The results of repeating the same test are shown in
Table~\ref{table:RandomTestFIPS}, which does not agree with the data
shown in Table~2 of \cite{Tong:ImageCipher:IVC07}.
\end{enumerate}

\begin{table}[htbp]
\centering \caption{Randomness test results of the chaotic compound
sequence generated from the key $(x_0, y_0)=(0.32145645647836,
0.48124356788345)$. For runs tests, the two output values are the
numbers of 0-bit and 1-bit runs,
respectively.}\label{table:RandomTestFIPS} \footnotesize
\begin{tabular}{*{4}{c|}c}\hline
\multicolumn{2}{c|}{Test item} & Required interval & Output value(s) & Result\\
\hline \hline \multicolumn{2}{c|}{Monobit test} & 9725 -- 10275 &
9968 & Pass\\ \hline \multirow{6}{*}{Runs test}
             & $r=1$    &  2315 -- 2685 & 2124, 2142 & Fail\\
\cline{2-5}  & $r=2$    &  1114 -- 1386 & 962, 966  & Fail\\
\cline{2-5}  & $r=3$    &  527 -- 723   & 537, 498  & Fail\\
\cline{2-5}  & $r=4$    &  240 -- 384   & 266, 273  & Pass\\
\cline{2-5}  & $r=5$    &  103 -- 209   & 153, 167  & Pass\\
\cline{2-5}  & $r\ge6$  &  103 -- 209   & 301, 297  & Fail\\
\cline{2-5}  & $r\ge 26$&  0 -- 0       & 3, 3    & Fail\\
\hline \multicolumn{2}{c|}{Poker test} &  2.16 -- 46.17 & 799.37 &
Fail\\ \hline
\end{tabular}
\end{table}

\color{black}

To investigate the level of randomness of the chaotic compound
sequence $\{z_k\}_{k=1}^{MN}$ generated by iterating
Eq.~(\ref{eq:compoundmap}), 100 binary sequences have been tested
for the encryption of $256\times 256$ images with the test suite
proposed in \cite{Rukhin:TestPRNG:NIST}. The secret keys to
generate the 100 binary sequences were chosen randomly. For each
test, the default significance level 0.01 was adopted. The results
are shown in Table~\ref{table:test}, from which one can see that the
compound chaotic function Eq.~(\ref{eq:compoundmap}) cannot be used
as a good random number generator.

\begin{table}[!htbp]
\centering\caption{The performed tests with respect to a significance
level 0.01 and the number of sequences passing each test in 100
randomly generated sequences.}\label{table:test}
\begin{tabular}{c|c}
\hline Name of Test & Number of Passed Sequences\\
\hline\hline Frequency  & 91\\
\hline Block Frequency ($m=100$) &  0\\
\hline Cumulative Sums-Forward & 88\\
\hline Runs             & 0 \\
\hline Rank             & 67\\
\hline Non-overlapping Template ($m=9$, $B=101001100$) & 48\\
\hline Serial ($m=16$) & 0\\
\hline Approximate Entropy ($m=10$)& 0\\
\hline FFT                         & 0\\ \hline
\end{tabular}
\end{table}

\subsection{Weak keys}

For the image encryption scheme under study, it is found that some
keys will cause some or even all encryption parts to fail, due to
the existence of some fixed points of the chaotic maps involved:
$f_0(1)=1$, $f_1(1)=1$, $f_1(0)=0$, $f_1(-1)=-1$. Four typical
classes of weak keys and the negative influences on the randomness
of the chaotic sequences are listed below:
\begin{enumerate}
\item \textit{$x_0=1$}: $f(x_0)=1$ $\Rightarrow$ $S_2(j)\equiv M-1$;
\item \textit{$y_0=1$}: $f_1(y_0)=1$, only $f_1(y)$ is iterated in
Eq.~\eqref{eq:compoundmap} $\Rightarrow$ $S_1(k)\equiv 255$,
$S_3(i)\equiv N-1$;
\item $y_0=-1$: $f_1(y_0)=-1$ $\Rightarrow$ $S_3(i)\equiv 0$;
\item \textit{$x_0\ge 0$, $y_0=0$}: $f_1(y_0)=0$, only $f_1(y)$ is
iterated in Eq.~\eqref{eq:compoundmap} $\Rightarrow$ $S_1(k)\equiv
128$, $S_3(i)\equiv N/2$.
\end{enumerate}
By combining the above conditions, three extremely weak keys can be
found from the above general ones:
\begin{itemize}
\item \textit{$x_0=1$, $y_0=1$}: $S_1(k)\equiv 255$, $S_2(j)\equiv M-1$,
$S_3(i)\equiv N-1$;
\item \textit{$x_0=1$, $y_0=-1$}: $S_1(k)\equiv 0$, $S_2(j)\equiv M-1$,
$S_3(i)\equiv 0$;
\item \textit{$x_0=1$, $y_0=0$}: $S_1(k)\equiv 128$, $S_2(j)\equiv M-1$,
$S_3(i)\equiv N/2$.
\end{itemize}
Furthermore, whenever $(x_{k_0},y_{k_1})$ satisfies one of the
above-listed conditions in the process of iterating
Eq.~\eqref{eq:compoundmap}, the corresponding secret key $(x_0,y_0)$
is also found to be weak. For instance, from $f_0(-1)=f_0(0)=1$,
$f_1(-0.5)=1$ and $f_1(0.5)=-1$, the following examples can be
derived easily: (1) $x_0\in\{0,-1\}$; (2) $y_0=-0.5$; (3) $y_0=0.5$.
From these examples, one can further discover some extremely weak
keys as follows:
\begin{itemize}
\item \textit{$x_0\in\{0,-1\}$, $y_0\in\{-0.5,1\}$}: $S_1(k)\equiv 255$,
$S_2(j)\equiv M-1$, $S_3(i)\equiv N-1$;
\item \textit{$x_0=0$, $y_0=0.5$}: $S_1(2)=255$, $S_1(k)\equiv 0$ for
$k\neq 2$, $S_2(j)\equiv M-1$, $S_3(i)\equiv 0$;
\item \textit{$x_0=0$, $y_0=-1$ or $x_0=-1$, $y_0\in\{-1,0.5\}$}:
$S_1(1)=255$, $S_1(k)\equiv 0$ for $k\geq 2$, $S_2(j)\equiv M-1$,
$S_3(i)\equiv 0$;
\item \textit{$x_0=0$, $y_0=0$}: $S_1(k)\equiv 128$, $S_2(j)\equiv M-1$,
$S_3(i)\equiv N/2$;
\item \textit{$x_0=-1$, $y_0=0$}: $S_1(1)=255$, $S_1(k)\equiv 128$ for
$k\geq 2$, $S_2(j)\equiv M-1$, $S_3(i)\equiv N/2$.
\end{itemize}

\subsection{Equivalent keys}

\begin{figure}[htbp]
\centering
\begin{minipage}[t]{\imgwidth}
\centering
\includegraphics[width=\textwidth]{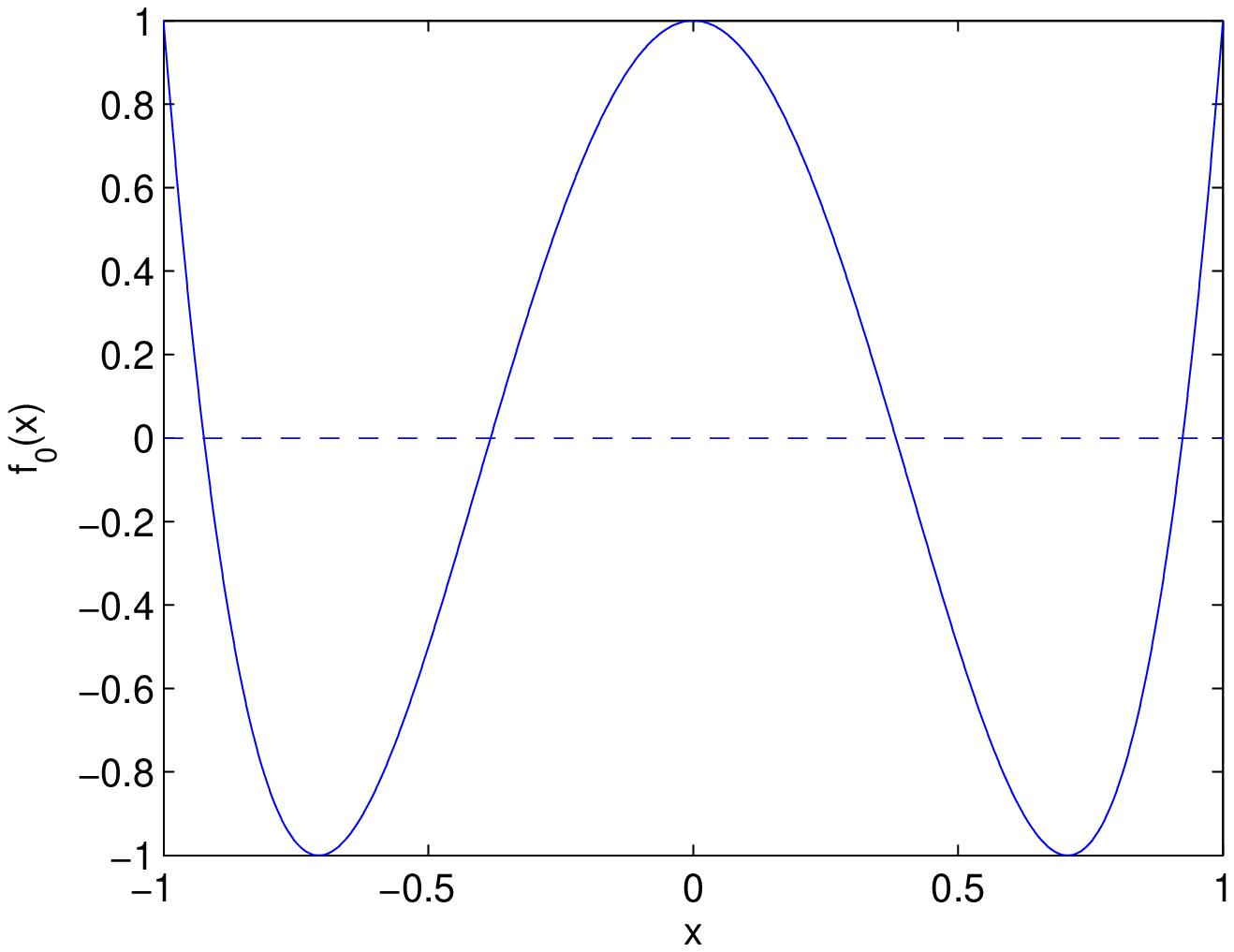}
a)
\end{minipage}
\begin{minipage}[t]{\imgwidth}
\centering
\includegraphics[width=\textwidth]{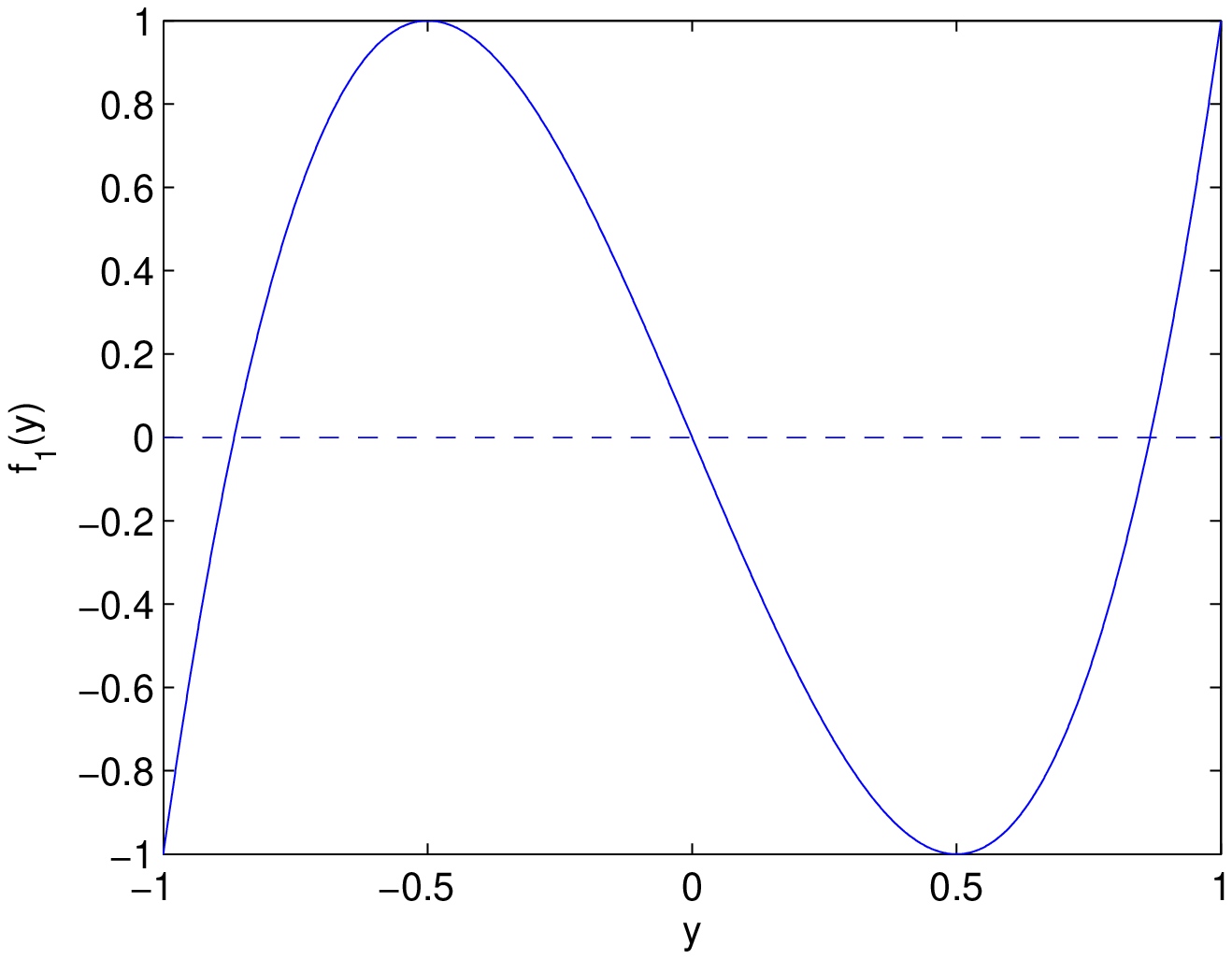}
b)
\end{minipage}
\caption{The images of functions $f_0(x)$ and $f_1(y)$}
\label{fig:ImageFunction}
\end{figure}

Equivalent keys mean some different keys that generate the same
cipher-image for any given plain-image, i.e., they are completely
equivalent to each other. From Fig.~\ref{fig:ImageFunction}a) one
can see that function $f_0$ may have four points whose functional
values are the same: $\pm x$, $\pm \sqrt{1-x^2}$. From
Fig.~\ref{fig:ImageFunction}b) one can see that function $f_1$ may
have three points whose functional values are the same: $y$,
$\frac{-y\pm\sqrt{3-3y^2}}{2}$.

Since only the field of rational number is considered, one can see
that $(x_0, y_0)$ and $(-x_0, y_0)$ are equivalent when $|y_0|\ge
|x_0|$.

\subsection{Low sensitivity to plaintext changes}

In \cite[Sec. 4.4]{Tong:ImageCipher:IVC07} the authors claim that
their scheme is sensitive to plaintext changes, which is, however,
not true. From Eq.~(\ref{eq:fullencryption}) one can easily see that
changing one bit of $I(i^*,j^*)$ influences the same bit of
$I'(i,j)$, only. Note that this low sensitivity is actually a common
problem with all XOR-based encryption systems. But it becomes
trivial if the key is not repeatedly used. In this case, it is rare
that two slightly different plaintexts are encrypted by the same
keystream.

\subsection{A remark on the compound chaotic map}

In Section~2.2 of \cite{Tong:ImageCipher:IVC07}, the authors have
provided some theoretical results about the compound chaotic map
defined as follows:
\begin{equation}\label{eq:compoundmap2}
F(x)=\begin{cases}%
8x^4-8x^2+1, & x<0\\
4x^3-3x, & x\geq 0,
\end{cases}
\end{equation}
and claimed that ``$F(x)$ can be employed as ideal sequence
cipher''. Unfortunately, as shown in Eq.~\eqref{eq:compoundmap},
what they actually employed in the design of the image encryption
scheme is a simple combination of two separately (but not
independently) iterated chaotic maps $f_0$ and $f_1$, which has
nothing to do with the above compound chaotic map
\eqref{eq:compoundmap2}. This makes all the theoretical results
given in \cite[Section~2.2]{Tong:ImageCipher:IVC07} completely
irrelevant to their image encryption scheme.

\section{Differential chosen-plaintext attack}\label{sec:Cryptanalysis}

In \cite[Sec. 4.6]{Tong:ImageCipher:IVC07} the authors claim that
their scheme can withstand chosen-plaintext attack efficiently. It
is found, however, that their scheme can be broken with only three
chosen plain-images.

The proposed attack is based on the following fact: given two
plain-images $\mymatrix{I_1}$, $\mymatrix{I}_2$ and the
corresponding cipher-images $\mymatrix{I}'_1$, $\mymatrix{I}'_2$,
one can easily verify that $I_1'(i,j)\oplus I_2'(i,j)=
I_1(i^*,j^*)\oplus I_2(i^*,j^*)$, where $j^*=(j-S_3(i))\bmod N$ and
$i^*=(i-S_2(j^*))\bmod M$. This means that the XOR substitution
operations disappear and only the permutations remain. According to
the quantitative cryptanalysis given in
\cite{Li:PermutationAnalysis:2004}, permutation-only ciphers are
always insecure against plaintext attacks, and only
$\lceil\log_{256}(MN)\rceil$ plain-images are required for a
successful chosen-plaintext attack. Once the permutation part is
broken, the XOR substitution can be cracked easily. This is a
typical \textit{divide-and-conquer} (DAC) attack that breaks
different encryption components separately.

Since the permutations in the image encryption scheme are a simple
combination of $N$ row-shift and $M$ column-shift operations, the
number of required differential plain-images will not be greater
than 2, even when $\lceil\log_{256}(MN)\rceil>2$. This means that
only 3 chosen plain-images suffice to implement the attack. In the
sequel, the DAC attack is described step by step.
\begin{itemize}
\item \textit{Breaking $\{S_3(i)\}_{i=1}^M$ (i.e., vertical shift
operations)}

If two plain-images $\mymatrix{I}_1$ and $\mymatrix{I}_2$ are chosen
such that each row of $\mymatrix{I}_1\oplus \mymatrix{I}_2$ contains
identical pixel values, then the horizontal circular shift
operations will be canceled and only vertical ones are left. If
further $\mymatrix{I}_1$ and $\mymatrix{I}_2$ are chosen such that
each column of $\mymatrix{I}_1\oplus \mymatrix{I}_2$ has an
unambiguous pattern to recognize the value $S_3(i)$, then the
vertical shift operations are broken. For example, one can choose
$\mymatrix{I}_1$ and $\mymatrix{I}_2$ as
\begin{equation}\label{eq:DifferentialImage}
I_1(:,j)\oplus I_2(:,j)=\begin{cases}%
0, & j=1,\\
255, & 2\leq j\leq N.
\end{cases}
\end{equation}
In this case, by looking for the new position of the sole black
pixel in each column, one can immediately derive all values of
$\{S_3(i)\}_{i=1}^M$.

\item \textit{Breaking $\{S_2(j)\}_{j=1}^N$ (i.e., horizontal shift operations)}

Once all vertical shift operations have been broken, one can use the
same strategy to break the horizontal shift operations. For this
purpose, one needs to choose $\mymatrix{I}_1$ and a new plain-image
$\mymatrix{I}_3$ such that each column of $\mymatrix{I}_1\oplus
\mymatrix{I}_3$ contains identical pixel values and each row has an
unambiguous pattern so as to recognize the value of $S_2(j)$. For
example, one can choose $\mymatrix{I}_1$ and $\mymatrix{I}_3$ as
\[
I_1(i,:)\oplus I_3(i,:)=\begin{cases}%
0, & i=1,\\
255, & 2\leq i\leq M.
\end{cases}
\]
In this case, by looking for the new position of the sole black
pixel in each row, one can immediately derive all values of
$\{S_2(j)\}_{j=1}^N$.

\item \textit{Breaking $\{S_1(i)\}_{i=1}^{MN}$ (i.e., XOR substitutions)}

After the values of $\{S_2(j)\}_{j=1}^N$ and $\{S_3(i)\}_{i=1}^M$
are obtained, the encryption scheme becomes a simple XOR-based
stream cipher, and $\{S_1(k)\}_{k=1}^{MN}$ can immediately be
recovered via
\[
S_1((j-1)\cdot M+i)=I_1(i,j)\oplus I_1'(i^*,j^*),
\]
where $i^*=(i+S_2(j))\bmod M$ and $j^*=(j-S_3(i^*))\bmod N$.
\end{itemize}

\begin{figure}[htbp]
\centering
\begin{minipage}[t]{0.5\imgwidth}
\includegraphics[width=\textwidth]{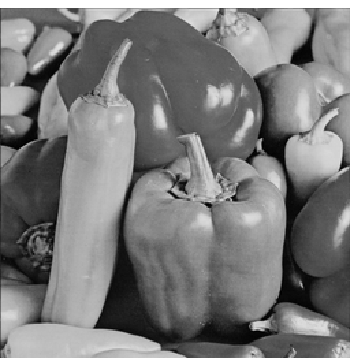}
a) $\mymatrix{I_1}$
\end{minipage}
\begin{minipage}[t]{0.5\imgwidth}
\includegraphics[width=\textwidth]{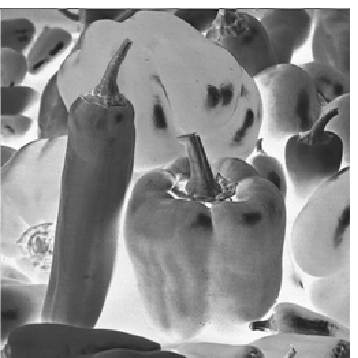}
b) $\mymatrix{I}_2$
\end{minipage}
\begin{minipage}[t]{0.5\imgwidth}
\includegraphics[width=\textwidth]{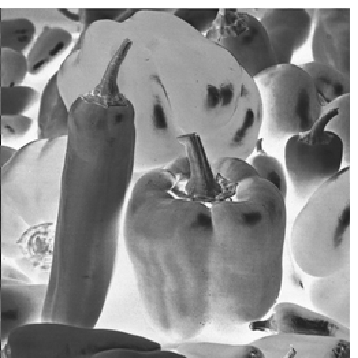}
c) $\mymatrix{I}_3$
\end{minipage}
\begin{minipage}[t]{0.5\imgwidth}
\includegraphics[width=\textwidth]{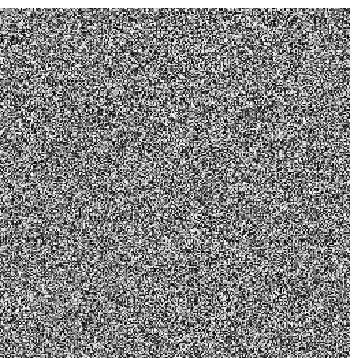}
d) $\mymatrix{I}'_4$
\end{minipage}\\
\begin{minipage}[t]{0.5\imgwidth}
\includegraphics[width=\textwidth]{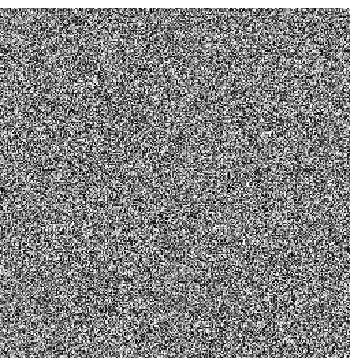}
e) $\mymatrix{I}'_1$
\end{minipage}
\begin{minipage}[t]{0.5\imgwidth}
\includegraphics[width=\textwidth]{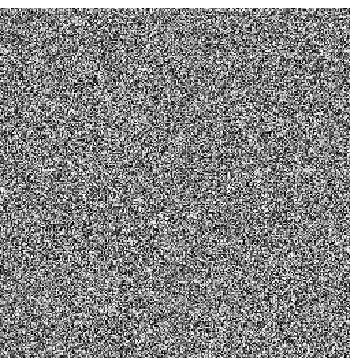}
f) $\mymatrix{I}'_2$
\end{minipage}
\begin{minipage}[t]{0.5\imgwidth}
\includegraphics[width=\textwidth]{ChosenPlainImage1_E}
g) $\mymatrix{I}'_3$
\end{minipage}
\begin{minipage}[t]{0.5\imgwidth}
\includegraphics[width=\textwidth]{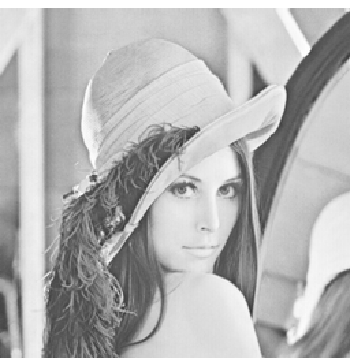}
h) $\mymatrix{I}_4$
\end{minipage}
\caption{The proposed differential chosen-plaintext attack: a
demonstration}\label{fig:DifferentialAttack}
\end{figure}

To validate the performance of the above attack, some experiments
have been carried out for some chosen plain-images of size
$256\times 256$. Here, the experimental results with the random
secret key used in Section~\ref{ssec:Randomness} are reported. One
plain-image ``Peppers'' is chosen as $\mymatrix{I}_1$, and the
second plain-image is chosen such that the differential image
$\mymatrix{I}_1\oplus \mymatrix{I}_2$ is as shown in
Eq.~(\ref{eq:DifferentialImage}). The third plain-image is chosen
such that $\mymatrix{I_1}\oplus\mymatrix{I_3}=(\mymatrix{I}_1\oplus
\mymatrix{I}_2)^T$. These three chosen plain-images and the
corresponding cipher-images are shown in
Fig.~\ref{fig:DifferentialAttack}. The recovered pseudo-random
sequences are used to decrypt a new cipher-image $\mymatrix{I}'_4$,
which is shown in Fig.~\ref{fig:DifferentialAttack}d), and the
result is given in Fig.~\ref{fig:DifferentialAttack}h).

\section{Conclusion}\label{sec:conclusion}

The security of a recently published image encryption scheme based
on a compound chaotic sequence has been studied. It is found that
the scheme can be broken with only three chosen plain-images. In
addition, it is found that the scheme has some weak keys and
equivalent keys, and that the scheme is not sufficiently sensitive
to the changes of plain-images. Furthermore, the pseudo-random
number sequence generated by iterating the compound chaotic function
is found not to be sufficiently random for secure encryption. In
summary, the scheme under study is not secure enough. Therefore, it
is not be recommended for applications requiring a high level of
security.

\begin{ack}
This research was supported by the City University of Hong Kong
under the SRG grant 7002134. In particular, Shujun Li was supported
by a research fellowship of the Alexander von Humboldt Foundation of
Germany.
\end{ack}

\bibliographystyle{elsart-num}
\bibliography{IVC2}

\end{document}